# Anomalous transport properties in Nb/Bi$_{1.95}$Sb$_{0.05}$Se$_3$ hybrid structure

E. P. Amaladass[*], Shilpam Sharma, A. V. Thanikai Arasu, R.Baskaran, Awadhesh Mani[*]
*Condensed Matter Physics Division, Materials Science Group,
Indira Gandhi Centre for Atomic Research, HBNI, Kalpakkam-603102, India,*

*Abstract-* We report the proximity induced anomalous transport behavior in a Nb/Bi$_{1.95}$Sb$_{0.05}$Se$_3$ hetero-structure. Mechanically Exfoliated single crystal of Bi$_{1.95}$Sb$_{0.05}$Se$_3$ topological insulator (TI) is partially covered with a 100 nm thick Niobium superconductor using DC magnetron sputtering by shadow masking technique. The magneto-transport (MR) measurements have been performed simultaneously on the TI sample with and without Nb top layer in the temperature (T) range of 3-8 K, and a magnetic field (B) up to ± 15 T. MR on TI region shows Subnikov-de Haas oscillation at fields > ±5 T. Anomalous linear change in resistance is observed in the field range of –4T < B < +4T at which Nb is superconducting. At 0 T field, the temperature dependence of resistance on the Nb covered region revealed a superconducting transition (T$_C$) at 8.2 K, whereas TI area showed similar T$_C$ with the absence of zero resistance states due to the additional resistance from superconductor(SC)/TI interface. Interestingly below the T$_C$ the R(T) measured on TI showed an enhancement in resistance for positive field and prominent fall in resistance for negative field direction. This indicates the directional dependent scattering of the Cooper pairs on the surface of the TI due to the superposition of spin singlet and triplet states in the superconductor and TI respectively.

## I. Introduction

A surge of research interest is seen in bismuth-based materials such as Bismuth Selenide (Bi$_2$Se$_3$), Bismuth Antimonide (Bi$_{1-x}$Sb$_x$), Bismuth Telluride (Bi$_2$Te$_3$) and their derivatives, the so-called three-dimensional (3D) topological insulators(TI) [1-7]. These materials host two-dimensional (2D) metallic surface states with Dirac-like band dispersion co-existing with an insulating bulk states. Due to strong spin-orbit coupling and topological nature of bands the spins of electron in these surface states are tightly coupled to their momentum and are protected by time reversal symmetry (TRS). Therefore they are immune to localization and backscattering and thereby are promising candidates for spintronics applications [8, 9]. Also, the interplay between the symmetry protected states in TI with symmetry broken quantum mechanical states such as superconductivity (SC) and ferromagnetism (FM) provides a fertile playground to explore new phenomena. The proximity of FM or SC with TI is believed to give rise to novel quantum mechanical phenomena such as anomalous quantum Hall effect (AQHE) and the realization of Majorana Fermions [10-12]. Previous reports on SC/TI have predominantly shown the resistance enhancement at T$_C$ while a few have reported on proximity-induced superconductivity in TI [13, 14]. To shed more light on the interaction and proximity effect in TI/SC heterostructures, we have studied the transport properties of Nb thin film sputter deposited on a mechanically exfoliated Bi$_{1.95}$Sb$_{0.05}$Se$_3$ TI single crystal.

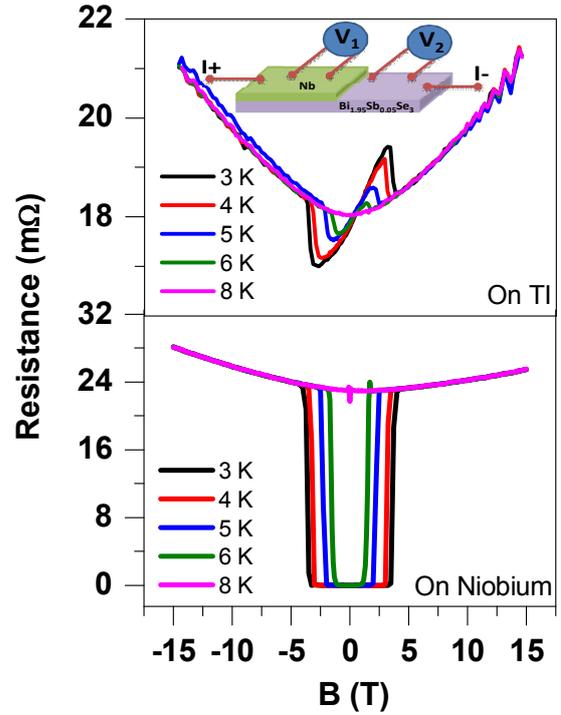

Fig 1. Magnetoresistance measured at different temperatures on TI region (top panel) and Nb/TI region (Bottom (panel). The inset shows the schematic of the sample and the contacts.

## II. Experimental Details

The single crystal of Bi$_{1.95}$Sb$_{0.05}$Se$_3$ was grown using high purity (99.999%) powders of Bi, Se, and Sb. Stoichiometric mixtures of elements are melted in an evacuated quartz tube at 850ºC for 24 hours followed by slow cooling at the rate of 2ºC/hour until 550ºC. After a halt at 550ºC temperature for 24 hours, the sample was cooled to room temperature. A part of rectangular shaped TI flakes was coated with 100 nm of Nb film in a magnetron sputtering system at a base pressure of ~ 10$^{-7}$ mbar. Temperature and magnetic field dependent resistance measurement were carried out in standard linear geometry using a commercial, 15 T Cryogen-free system from Cryogenic Ltd, UK. Resistance was measured by sourcing a constant current and measuring the voltage developed across the two leads from two different locations as schematically shown in the inset of Fig.

*III. RESULTS AND DISCUSSION*

The top panel in Fig. 1 shows the magnetoresistance measured on the TI at different temperatures, and in magnetic field up to ± 15 T applied perpendicular to the sample plane. At B > ±5 T the sample shows Subnikov-de Haas oscillations (SdH). From Fast Fourier Transform (FFT) analysis the frequency of oscillation is found to be 165 T. The absence of SdH oscillations for magnetic field parallel (not shown) to the sample plane shows that these oscillations are stemming from the 2D surface states in TI. Interestingly, on decreasing the magnetic field we see a sharp rise in the resistance at B < 4 T, and it peaks at ~ 3.25 T. Upon further reducing the field, the resistance falls linearly till –3.25 T and at B > -4 T it increases and merges with the quadratic MR behavior. Also, the magnitude of the linear change in R increases with decreasing temperature.

The bottom panel in Fig 1 shows the MR measured on the Nb covered region of TI. At 3 K and higher field it shows no SdH oscillations but at B < 4 T it shows a sharp dip and a zero resistance state in the field range of -4 T ≤ B ≤ 4 T. On further increasing the temperature to 8 K, MR shows similar dip but at the reduced field. These drop in MR is attributed to the temperature and field dependent critical field behavior of the Nb superconducting film on the TI. On comparing the two MR in Fig. 1 it is clear that the anomalous linear MR observed on TI region is precisely when the Nb is in the superconducting state. To get more insight into the observed MR on TI area while Nb in its superconducting state, R(T) was measured at different fields applied perpendicular to the sample plane. Fig. 2 shows the resistance versus temperature plots for the TI regions at different positive and negative field direction. Inset in the figure shows the R(T) measured on the Nb covered region of TI. $T_C$ (8.2 K) of Nb decreases with increasing magnetic field irrespective of the magnetic field direction. Interestingly the resistance measured on TI at T < $T_C$ at negative magnetic field, decreases as temperature decreases whereas for positive field it increases as temperature decreases. This is also inconsistent with the MR behavior where we see a linear growth in resistance for positive field direction and vice versa. The scattering of the spin singlet Cooper pairs from the spin-polarized or spin triplet TI is believed to cause the resistance enhancement below $T_C$ [13]. However, the magnetic field dependent resistance enhancement or diminution below $T_C$ of the superconductor clearly shows a different picture of the scattering mechanism. The superposition of the spin singlet state on the TI with the spin triplet states on superconductors and their magnetic field dependence as reported in [15] can lead to such changes in MR. However, further experiments like dynamic conductance measurements can shed more light on these observations.

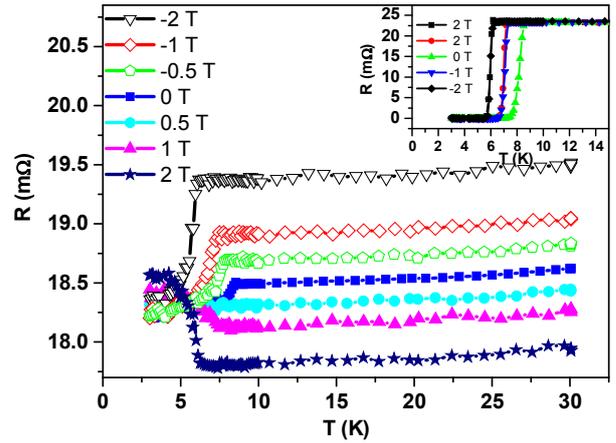

Fig 2. Temperature dependence of resistance measured on TI region of the sample at different magnetic fields. The inset shows the R(T) measured on the Nb covered regions of the TI sample.

ACKNOWLEDGMENT

The authors gratefully acknowledge UGC-DAE-CSR Kalpakkam node for providing access to the 15 T cryogen free magnetoresistance measurement system.

*Email: edward@igcar.gov.in; mani@igcar.gov.in*